\documentclass[twocolumn]{article}
\usepackage[a4paper, total={180mm, 243mm}]{geometry}
\usepackage[utf8]{inputenc}
\usepackage{lipsum}
\usepackage{graphicx}
\usepackage{siunitx}
\usepackage[version=4]{mhchem} 
\usepackage{titling}
\usepackage{circledsteps}
\usepackage{subcaption}

\DeclareCaptionLabelSeparator{custom}{\textbar}
\DeclareCaptionFormat{custom}{\textsf{\textbf{#1 #2} \small #3}}
\graphicspath{{./figures}} 

\makeatletter
\newcommand{\showfontsize}{\f@size{} pt}
\makeatother

\usepackage[
    backend=biber,
    style=nature,
    url=false,
    isbn=false,
    date=year
]{biblatex}

\newcommand{\reffig}[2]{Fig.~\ref{#1}{#2}}

\title{Soliton Pulses in Photonic Crystal Fabry-Perot Microresonators}

\author{Thibault Wildi$^1$, 
        Mahmoud A. Gaafar$^1$,
        Thibault Voumard$^1$,
        Markus Ludwig$^1$,
        Tobias Herr$^{1,2,*}$}
\date{%
    \small $^1$Deutsches Elektronen-Synchrotron DESY, Notkestr. 85, 22607 Hamburg, Germany \\
    \small $^2$Physics Department, Universität Hamburg UHH, Luruper Chaussee 149, 22761 Hamburg, Germany\\
    \small $^*$tobias.herr@desy.de
}

\begin{document}

\maketitle

\textbf{
Dissipative Kerr solitons (DKSs) in high-Q microresonators enable applications in sensing, communication, and signal processing. Until now, DKSs driven by continuous-wave (CW) lasers are exclusively generated in ring-type resonators. Complementary to ring-type resonators, Fabry-Perot resonators could enable new approaches to dispersion engineering, addressing a key challenge of DKS technology. However, DKS generation in a CW-driven Fabry-Perot microresonator has not yet been achieved. Here, we demonstrate for the first time CW-driven DKSs in a high-Q Fabry-Perot microresonator. Fabricated in a wafer-level process, two photonic crystal reflectors in a waveguide form the chip-integrated resonator and define its dispersion. The intrinsic \mbox{Q-factor} of \mbox{4~million} is propagation-loss limited.
In principle, each cell of the photonic crystal reflector can be tailored, opening a design space beyond traditional dispersion engineering, with potential for future extension of DKSs to visible and other currently inaccessible wavelengths.
Beyond DKSs, this creates opportunities for filter-driven pulse formation, engineered spectra and broadband phase-matching in microresonators.
}

Dissipative Kerr solitons (DKSs) \cite{ leo:2010,herr:2014a,kippenberg:2018, gaeta:2019} in laser-driven dielectric microresonators provide access to metrology-grade femtosecond sources and broadband frequency combs with repetition rates from tens of GHz to multiple THz. 
They are self-enforcing solutions to the Lugiato-Lefever equation (LLE) and can emerge in \textit{high-quality factor} (Q) microresonators from the balance between \textit{anomalous} group delay dispersion (GDD), loss and Kerr-nonlinearity under (typically) continuous-wave (CW) laser driving.
Intriguing nonlinear dynamics including soliton crystals \cite{cole:2017}, soliton molecules \cite{weng:2020}, synchronization between resonators \cite{jang:2018}, and discrete photonic time crystals \cite{taheri:2022} have been observed. 
DKSs also have demonstrated potential in cross-disciplinary applications; examples include data transmission \cite{marin:2017} and processing \cite{tan:2021}, ranging \cite{trocha:2018, suh:2018}, microwave photonics \cite{lucas:2020}, dual-comb spectroscopy \cite{suh:2016}, and astronomical spectrograph calibration \cite{obrzud:2019,suh:2019}. Combined with potential for wafer-level fabrication this positions DKSs as a transformative technology for mobile and space applications ranging from communication and navigation to ultrafast sensing and environmental monitoring.

To this day, DKSs have almost exclusively been pursued in traveling-wave ring-type resonators (incl. whispering-gallery-mode and race-track resonators), as they can offer high-Q and broadband control over dispersion: indeed through specific combinations of suitable materials and waveguide geometry, dispersion can be \emph{engineered} in certain wavelength intervals in order to achieve the required anomalous GDD. In this regard, integrated ring-type microresonators stand out as they provide simple yet effective ‘tuning knobs’ (such as waveguide height and width) that can be used to adjust a resonator's dispersion. 
In addition, advanced techniques for narrowband dispersion modification in high-Q rings have been demonstrated including shifting of resonance frequencies via mode-coupling between cross-polarized modes \cite{ramelow:2014}, fundamental and higher-order transverse modes \cite{li:2018}, counter-propagating modes \cite{yu:2021}, as well as mode-hybridization in concentric resonators \cite{kim:2017} and between modes in distinct resonators \cite{helgason:2021, tikan:2021}, which can lead to easier initiation of DKSs, higher-power efficiency, and novel nonlinear phenomena. However, these effects are limited in their strength and spectral range, rendering engineering of DKS spectral and temporal shape difficult and access to new wavelength domains especially challenging. More generally, this is related to the challenge of phase-matching, which also impacts other approaches to microresonator frequency combs via cascaded four-wave mixing \cite{delhaye:2007} and switching waves \cite{lobanov:2015, xue:2015, parra-rivas:2016}. Many advances could be enabled by additional degrees of freedom in microresonator design, with strong influence on dispersion.

\begin{figure*}[t]
    \centering
    \includegraphics[width=\textwidth]{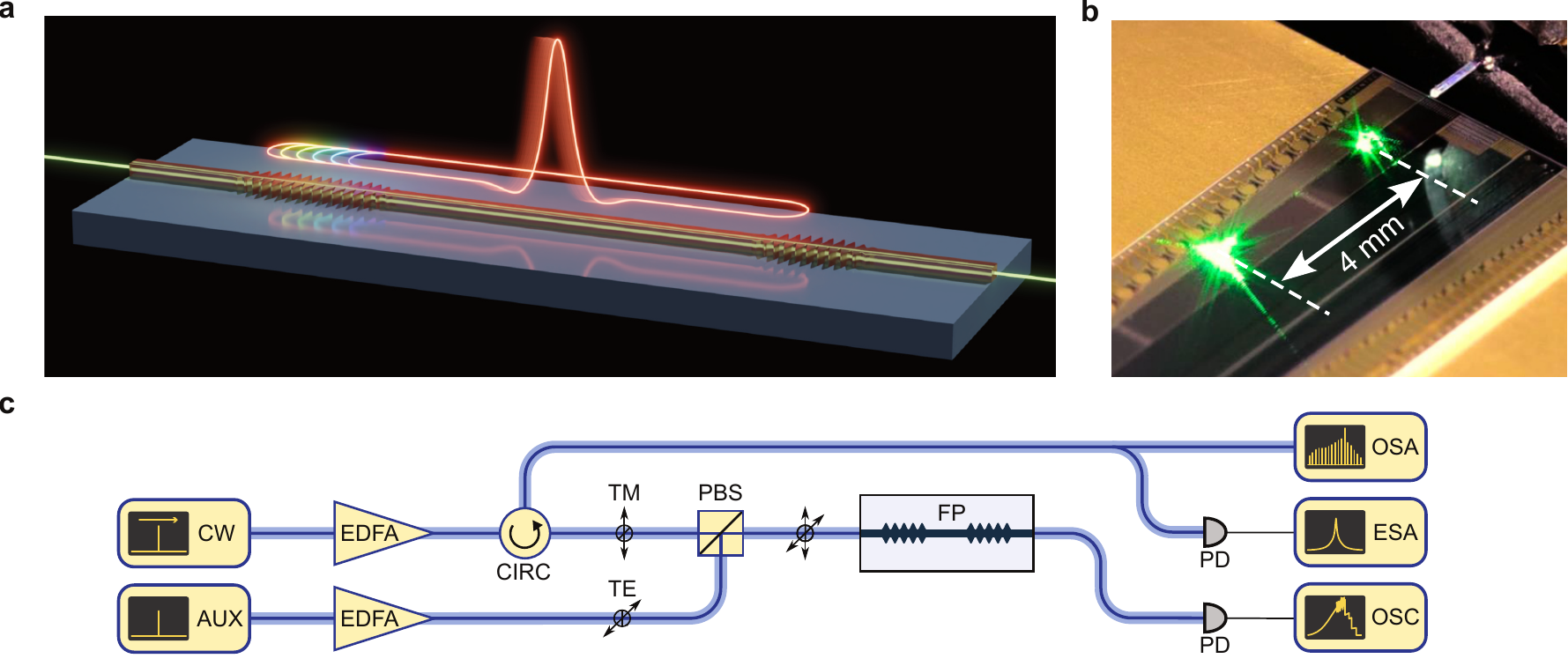}
    \caption{
        \textbf{Soliton generation in chip-integrated photonic crystal Fabry-Perot cavities.} 
        \textbf{\textsf{a}},~Illustration of a dissipative Kerr soliton (DKS) under continuous-wave pumping inside a chip-integrated photonic crystal Fabry-Perot (FP) microresonator with reflector-induced anomalous dispersion.
        \textbf{\textsf{b}},~Photograph of a CW-driven FP microresonator.  Scattered coincidental third-harmonic generation indicates the position of the PCRs.
        \textbf{\textsf{c}},~Setup for exciting DKSs inside the FP microresonator. A transverse magnetic (TM) polarized continuous-wave (CW) pump laser is tuned into resonance while the resonator is thermally stabilized by an auxiliary continuous-wave laser (AUX) in the transverse electric (TE) mode. EDFA: erbium-doped fiber amplifier; CIRC: circulator; PBS: polarizing beam splitter; PD: photodetector; OSC: oscilloscope; ESA: electrical spectrum analyzer; OSA: optical spectrum analyzer.
        }
    \label{fig:1}
\end{figure*}
A similar need for additional degrees freedom in designing table-top laser cavities was initially addressed, almost three decades ago, when high-reflectivity Bragg-mirrors were introduced for modification and tailoring of dispersion in ultrafast laser cavities \cite{szipoecs:1994, kaertner:1997}. Inspired by their success in ultrafast lasers, Bragg-mirrors have also been applied to short fiber-based Kerr-nonlinear microresonators resulting in four-wave-mixing and stimulated Brillouin scattering \cite{braje:2009}, as well as the first demonstration of pulse-driven DKS \cite{obrzud:2017}. Advances in photonic integration recently led to the first integrated Fabry-Perot (FP) resonator with photonic crystal mirrors for comb generation \cite{yu:2019} as well as microresonators with disperison engineered reflecting structures, based on inverse design \cite{lucas:2022, ahn:2022} and Fourier-synthesis \cite{moille:2022}. 
Despite these major advances, a DKS-supporting platform whose broadband dispersion is not limited to effective waveguide or whispering-gallery mode dispersion has not yet been demonstrated.

Here, we demonstrate for the first time CW-driven DKSs in a standing-wave FP microresonator. The resonator is implemented on-chip and composed of two photonic crystal reflectors (PCRs) in a waveguide (\reffig{fig:1}{a}).
Significantly, across its entire bandwidth the dispersion of the resonator is dominated by the PCRs (not the waveguide), demonstrating new opportunities for dispersion engineering in a system capable of supporting DKSs. Further, the resonator's intrinsic Q-factor, albeit not a record-high for a wafer-scale process \cite{liu:2021, ji:2021}, is on-par with equivalent ring resonators fabricated in the same commercial platform. This establishes integrated FP resonators as a powerful complement to ring resonators.
\begin{figure*}[t!]
    \centering
    \includegraphics[width=\textwidth]{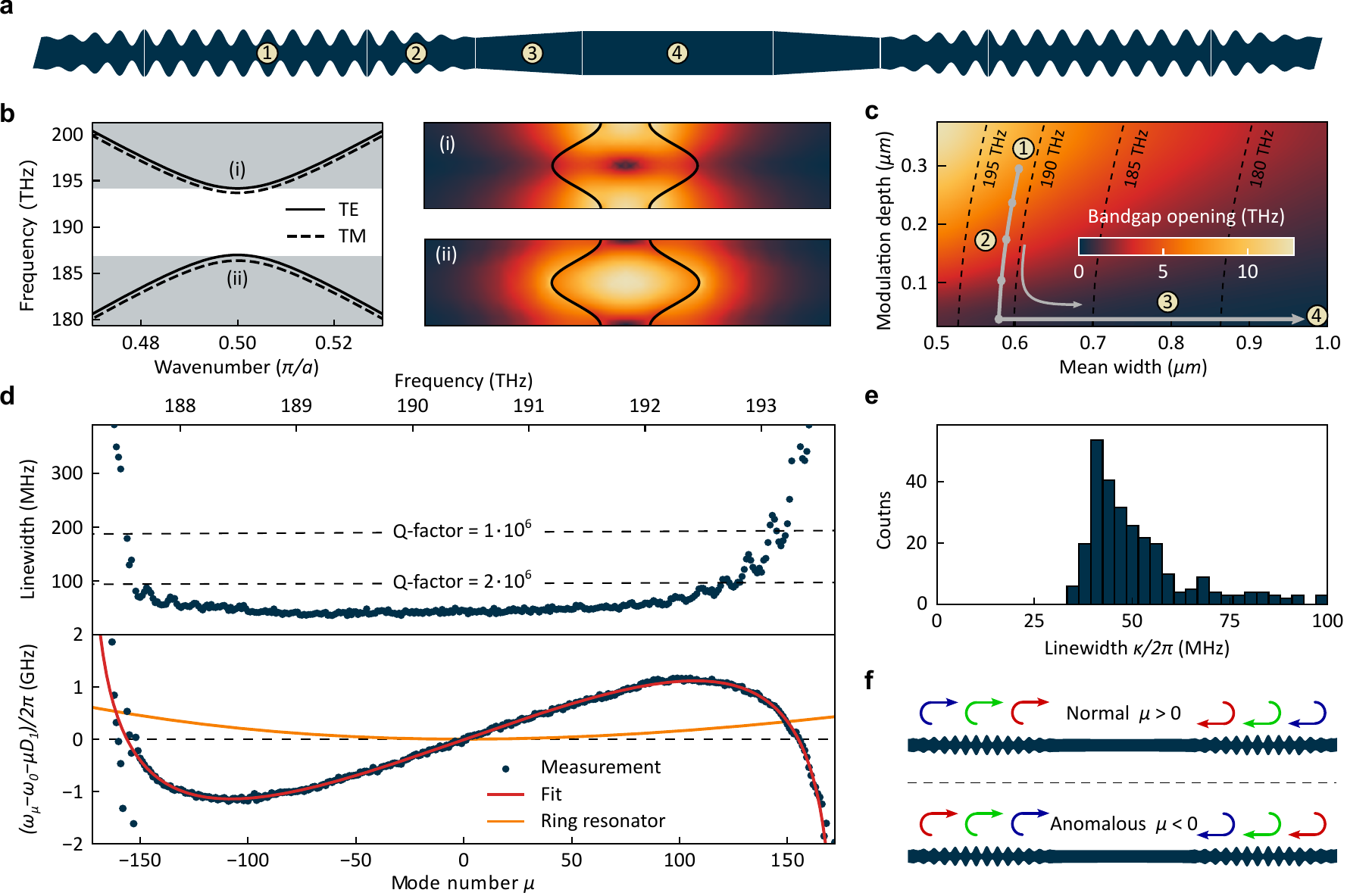}
    \caption{
        \textbf{Design of photonic crystal Fabry-Perot cavities.} 
        \textbf{\textsf{a}},~Schematic of a photonic crystal Fabry-Perot (FP) microresonator (not to scale), composed of photonic crystal reflectors (PCRs) \Circled{1}, adiabatic tapers \Circled{2}, linear tapers \Circled{3} and intracavity waveguide \Circled{4}. 
        \textbf{\textsf{b}},~Left panel: Photonic band diagram for transverse electric (TE) and transverse magnetic (TM) modes. The wavenumber is given in units of $\pi/a$ where $a$ is the unit cell length. The bandgap, i.e. the reflection bandwidth is highlighted in white. Right panels: TE mode profiles (electric field strength) of upper and lower frequency modes of an example PCR unit cell with the corresponding cell contour overlaid in black.
        \textbf{\textsf{c}},~Map of the bandgap opening and center frequency as function of corrugation depth and mean width of the lattice cell for a fixed PCR period (i.e. cell length) of \SI{475}{\nm}. The PCR can be described by its trajectory through this design space (where the third dimension, the cell length, is not shown).
        \textbf{\textsf{d}},~Upper panel: Measured resonance linewidths as function of relative mode number $\mu$ (cf. main text). Lower panel: Measured dispersion (cf. main text) of a photonic crystal FP resonator with an mean free-spectral range $\overline{\mathrm{FSR}}=D_1/2\pi = \SI{18.55}{\GHz}$. For comparison, the dispersion of an equivalent ring resonator with same $\overline{\mathrm{FSR}}$ and waveguide cross-section identical to the FP's intracavity waveguide is plotted in orange, corresponding to the intracavity waveguide contribution to the FP resonator's total dispersion. A fit based on a coupled-mode description of the PCRs is shown in red (cf. Methods).
        \textbf{\textsf{e}},~Histogram of the intrinsic resonance linewidths from \textbf{\textsf{d}} with a median value of \SI{47}{\MHz} corresponding to an intrinsic Q-factor of 4.0~million.
        \textbf{\textsf{f}},~Normal (anomalous) round-trip GDD corresponds to an increasing (decreasing) effective resonator length with increasing frequency. 
    }
    \label{fig:2}
\end{figure*}
\subsection*{Results}
The FP resonators are fabricated on-chip in a \SI{800}{\nm} thick silicon nitride layer and embedded in a fused silica cladding. This approach permits direct comparison with ring resonators routinely fabricated in the same material platform.
The two PCRs are implemented as submicron-scale sinusoidal corrugations in a waveguide. Each corrugation period corresponds to a unit cell in the PCR, which is characterized by its length, mean width, and corrugation depth (\reffig{fig:2}{a} \Circled{1}). The periodic corrugation induces a photonic bandgap that defines the PCR's reflection bandwidth (\reffig{fig:2}{b}); the length of the PCRs ($\sim$100~units cell per reflector) defines the reflectivity and thus the coupling strength to the waveguide that extends beyond the resonator to the chip's facets for light coupling. To design the PCRs, we first map the unit cell parameters to the photonic bandgap's opening and central frequency (\reffig{fig:2}{c}), which permits choosing the desired parameters. In the present case, we choose a constant unit cell for the main region of the PCR to create a bandgap centered around \SI{1570}{\nm}, the middle wavelength of our tunable CW laser. Note, that in principle each unit cell could have a different set of defining parameters, creating a large design space and permitting to craft highly-customized resonators.
An adiabatic taper (\reffig{fig:2}{a} \Circled{2}) connects the PCRs on both sides to an uncorrugated waveguide and suppresses losses due to the overlap mismatch between the fundamental guided mode of the uncorrugated waveguide and the PCR's fundamental Bloch mode \cite{palamaru:2001, lalanne:2003}. Transitioning from the PCR to the uncorrugated waveguide proceeds by gradually reducing the corrugation depth while simultaneously adjusting the mean width such that the bandgap center frequency is kept constant \cite{sauvan:2005}. Finally, upon reaching zero-corrugation, the waveguide linearly tapers up to a width of \SI{1.6}{\um} over length of \SI{200}{\um} to reduce loss from sidewall-roughness scattering while still keeping a strong mode-confinement (\reffig{fig:2}{a} \Circled{3}). A \SI{3400}{\um} long waveguide section forms the resonator cavity between the 2 PCRs (\reffig{fig:2}{a} \Circled{4}), defining the resonator's free-spectral range (FSR), in this case \SI{\sim19}{\GHz} which is within the K-frequency band and directly detectable by a photodiode and a microwave spectrum analyzer.

\begin{figure*}[t]
    \centering
    \includegraphics[width=\textwidth]{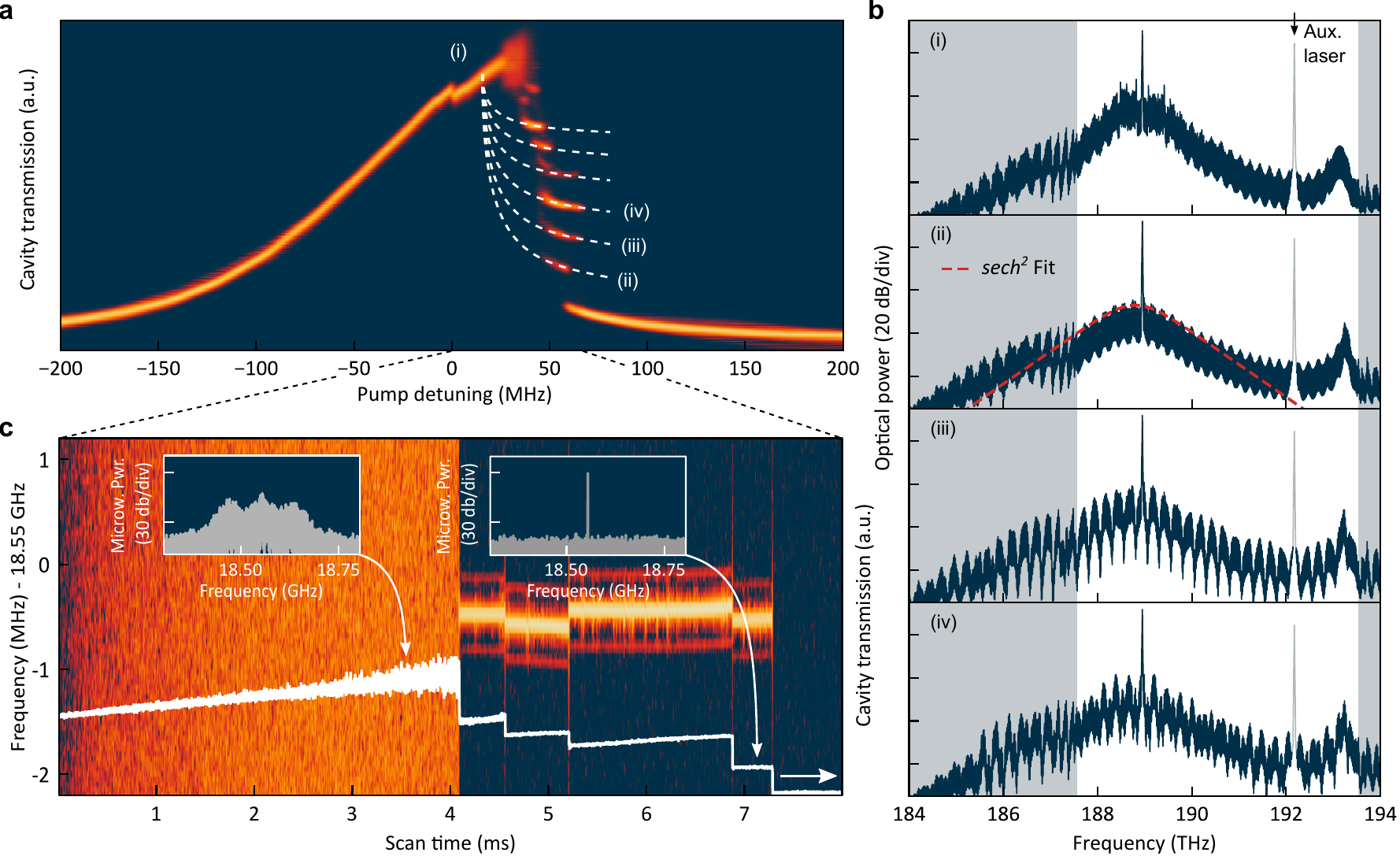}
    \caption{
        \textbf{Dissipative Kerr soliton in a photonic crystal Fabry-Perot resonator.}
        \textbf{\textsf{a}},~Heat-map compiled from a 100 transmission traces, highlighting multiple possible intracavity power evolutions as a function of pump detuning. The step-like features present at \SI{\sim50}{\MHz} detuning indicate quantized power levels, a canonical signature of DKS formation.
        \textbf{\textsf{b}},~Optical spectra of modulation instability (i), single (ii), 2- (iii) and 3- (iv) soliton states. The spectra are subject to a periodic modulation of their envelope due to reflection between the PCR and chip facet (cf. Methods). A dispersive wave is present on the blue-side of the spectra, due to the local presence of strong normal dispersion. The auxiliary laser's line (cf. Methods), is grayed out for clarity; the nominal PCR bandwidth is highlighted in white. A $sech^2$ envelope is fitted to the single soliton spectrum (i), corresponding to a transform-limited pulse duration of \SI{\sim300}{\fs}. 
        \textbf{\textsf{c}},~Spectrogram showing the evolution of the mirowave beatnote signal ($\SI{\sim18.5}{\GHz}$) as a function of the pump detuning. Overlayed in white is the corresponding resonator transmission signal. The insets show the beatnote in the modulation instability and soliton regime respectively (resolution bandwidth: \SI{100}{\kHz}).
        }
    \label{fig:3}
\end{figure*}

To characterize the fabricated resonators, we first measure the intrinsic linewidths in a strongly undercoupled resonator (\reffig{fig:2}{d}); undercoupling assesses the intrinsic cavity loss and permits direct comparison to (equally undercoupled) ring-type resonators fabricated on the same platform. The PCR unit cell of this resonator is designed to have a period of \SI{475}{\nm}, width of \SI{600}{\nm} and corrugation depth of \SI{300}{\nm}. Owing to the narrow width of the unit cell, the PCR only supports the fundamental modes, which is desirable for reproducible resonator characteristics. 
The resonator's spectrum covers over 300 longitudinal modes and an intrinsic linewidth of \SI{100}{\MHz} or below is maintained over a bandwidth of \SI{5}{\THz} (\SI{40}{\nm}).
The median intrinsic (undercoupled) linewidth over the mirror bandwidth (here defined as the spectral interval where the linewidth is consistently below \SI{100}{\MHz}) is \SI{47}{\MHz} (\reffig{fig:2}{e}). This corresponds to a median intrinsic Q-factor of $4.0 \cdot 10^6$, which is only limited by the \SI{\sim0.1}{\dB\per\cm} propagation loss inherent to the commercial platform used to fabricate the samples (cf. Methods), and thus on-par with ring-resonators fabricated through this commercial wafer-scale process.

Next, we measure the resonator's dispersion as shown in \reffig{fig:2}{d} in terms of the integrated dispersion $D_\mathrm{int} = \omega_\mu -\omega_0 - \mu D_1$, which quantifies the deviation of the resonance frequencies $\omega_\mu$ from a dispersion-free equidistant frequency grid as a function of the relative mode-number $\mu$. In this representation, anomalous (normal) dispersion appears as a convex (concave) curve. 
The strength of the local dispersion at a specific frequency can be estimated by choosing $D_1/(2\pi)$ to be the local FSR and expand $D_\mathrm{int} \approx \tfrac{1}{2}\mu^2 D_2$ around this frequency. Anomalous (normal) dispersion is then indicated by positive (negative) $D_2$. Here, $\mu$ is chosen such that $\mu=0$ coincides with the center of the reflector's bandwidth and $D_1/(2\pi)$ is chosen to be approximately the average $\overline{\mathrm{FSR}}$ over the reflector's bandwidth. For comparison, \reffig{fig:2}{d} also shows the dispersion curve of an equivalent ring-type resonator (with the same $\overline{\mathrm{FSR}}$ and the same waveguide cross-section as the FP intracavity waveguide), corresponding to the intracavity waveguide contribution to the FP resonator's total dispersion; the marked difference between both dispersion curves shows the dominating impact of the PCRs on the FP resonator’s dispersion.
The exact contribution of the PCRs to the dispersion can be calculated through their complex reflection coefficient (cf. Methods), which matches well the observations (also indicated in \reffig{fig:2}{d}). With the current PCR design, the resonator provides both normal and anomalous dispersion regimes, independently of the anomalous background contribution from the intracavity waveguide. This may be understood as a wavelength dependent effective reflection depth in the PCRs (\reffig{fig:2}{f}). Both dispersion regimes feature high-Q resonances and extend over an optical bandwidth that can support ultrashort femtosecond pulses. These results establish PCR-based dispersion modification as a powerful complement to waveguide-dispersion engineering that is compatible with high-Q factors.

Moreover, the measured dispersion of the resonance frequencies in \reffig{fig:2}{d} is free of strong local deviations such as avoided mode crossings (AMXs) which can arise from coupling between frequency degenerate counter-propagating modes or coupling between different transverse mode families. Here, these two mechanisms are absent: in contrast to rings, the former is not present in the FP configuration, and the latter is efficiently suppressed by the single-mode nature of the PCRs. The resulting smooth anomalous dispersion in conjunction with the high-Q provides favorable conditions for DKS formation \cite{herr:2014b} and enable controlled and reproducible resonator behavior. 

\begin{figure}[t]
    \centering
    \includegraphics[width=\columnwidth]{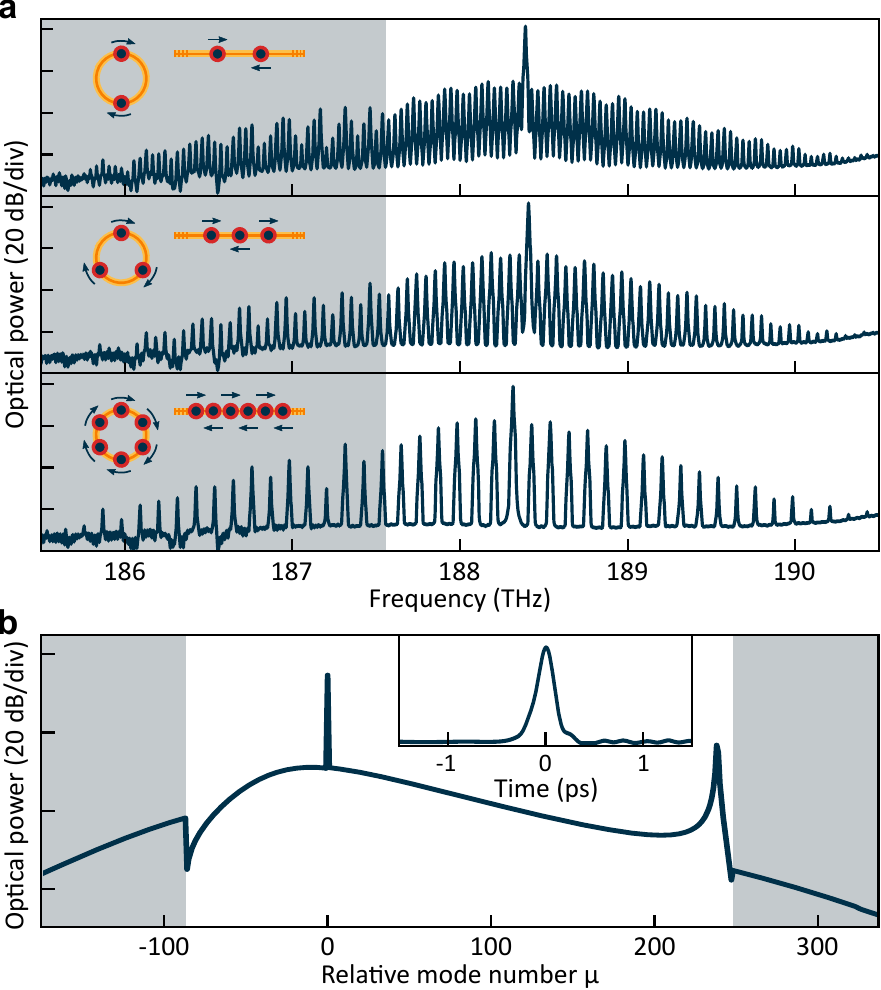}
    \caption{
        \textbf{\textsf{a}},~Recorded spectra of 2, 3 and 6-soliton crystal states. The PCR's nominal bandwidth is highlighted by a white background. A snap-shot crystalline distribution of DKSs in a linear FP cavity is shown as an inset along with an equivalent ring configuration.
        \textbf{\textsf{b}},~Simulated soliton spectra and corresponding temporal pulse profile (inset). The nominal PCR bandwidth is highlighted in white (note that here the relative mode number $\mu$ is defined with respect to the pumped mode).
    }
    \label{fig:4}
\end{figure}

To generate DKSs, a more strongly coupled resonator with a 15\% shorter input PCR (but with otherwise identical geometry) and a \SI{56}{\MHz} median total linewidth is used for more efficient operation.
The setup for this experiment is shown in \reffig{fig:1}{b}: The resonator is pumped using a tunable laser in the transverse-magnetic (TM) polarization in the anomalous dispersion regime (local $D_2/(2\pi) = \SI{210}{\kHz}$) at a wavelength of \SI{1587}{\nm} (\SI{188.9}{\THz}). 
Repeatedly scanning the pump laser (\SI{150}{\mW} on-chip) from blue- to red-detuned across a resonance, we record the resonator transmission signals and superpose them in \reffig{fig:3}{a}.
After an initial modulation instability (MI) state (i), characteristic step-features (ii-iv etc.) are visible, which are indicative of DKS formation \cite{herr:2014a}; each step corresponds to a certain numbers of solitons in the resonator that can be generated in a laser scan.
The corresponding MI spectrum, the comb spectrum of a single soliton with its characteristic sech$^2$-envelope, as well as more structured comb spectra of multiple-soliton states are shown in \reffig{fig:3}{b} where the white background highlights the nominal PCR bandwidth.
For clarity we note that all spectra exhibit a periodic modulation of their envelope caused by reflections between PCR and chip facet (cf. Methods); although of no concern here, this could be avoided by anti-reflection coating the chip's facet, index-matching fluid or evanescent waveguide coupling if necessary. From fitting the single DKS's spectrum with a $sech^2$ envelope (\reffig{fig:3}{b}~(ii)) we estimate the transform-limited soliton pulse duration to be \SI{\sim 300}{\fs}. Close to the blue-edge of the resonator's bandwidth, we observe a dispersive wave \cite{brasch:2016} as expected from the strongly normal dispersion in this spectral region. Due to a relaxed phase-matching condition outside the high-reflectivity bandwidth, all spectra extend beyond the nominal PCR bandwidth. 
To further confirm DKS generation, we record the microwave pulse repetition rate beatnote as a function of detuning (\reffig{fig:3}{c}). As expected, the beatnote transitions from a high-noise to a narrow-linewidth signal, when entering into a DKS state (each discrete transition to a different number of DKSs slightly changes the repetition rate). 
Finally, we also observe the formation of soliton crystals \cite{cole:2017} (\reffig{fig:4}{a}), a state commonly observed in traveling-wave microresonators where a self-organized train of equidistantly spaced soliton pulses circulates inside the resonator, (effectively increasing the pulse repetition rate and hence the comb line spacing by an integer factor). These observations further establish the standing-wave FP resonator as a DKS platform and strengthen the link to travelling-wave ring-type resonators.

Indeed, the nonlinear dynamics of standing wave-resonators is similar to the travelling-wave case, with an additional phase-shift due to cross-phase modulation between counter-propagating waves proportional to the average intracavity power \cite{obrzud:2017, cole:2018}; this effect can be compensated by a slight change in the pump laser detuning. Different from a conventional ring-type resonator, the dispersion in the present resonator includes localized and distributed contributions from the PCRs and the connecting waveguide, respectively (cf. Methods). Here, as the impact of both a single reflection off of a PCR or a single propagation through the waveguide have negligible effect on the DKS pulse, they can be lumped together. This implies that established mean-field models for ring resonators can also be used for describing the present FP resonators, with the inclusion of the aforementioned phase-shift. 

Complementing the experimental study of DKS formation in the new resonator platform we perform numeric simulations based on the frequency-domain formulation of the LLE \cite{chembo:2010, hansson:2014}, which readily permits inclusion of the spectrally-dependent measured dispersion and linewidths. The simulated spectrum and pulse shape of a single soliton state are shown in \reffig{fig:4}{b}. The simulation reproduces all spectral features (compared to the general envelope in \reffig{fig:3}{b} (ii-iv) and \reffig{fig:4}{a}) including the dispersive wave and the spectral extension beyond the nominal PCR bandwidth. The simulated pulse shows only minor deviations from DKSs generated in resonators with unlimited spectral support, corroborating that the reflectors' bandwidth is sufficient to support femtosecond DKS pulses.
\subsection*{Discussion}
In summary, we have shown for the first time DKS formation in a CW-driven standing-wave FP resonator. The chip-integrated resonator is composed of a waveguide that defines the resonator length and PCR mirrors that dominate and effectively define the resonator's dispersion. Through careful design, we achieve a high intrinsic Q-factor of 4~million, matching that of traveling wave type resonators fabricated by the same wafer-scale commercial foundry. We observe single, multiple (hence colliding) DKS pulses, as well as soliton crystals, highlighting the similarities to ring-type resonators.
Complementing established methods in travelling-wave type resonators, the FP geometry with PCR mirrors offers promising new design options for tailoring dispersion, achieving spectrally dependent Q-factors or broadband phase-matching, while allowing for exceptionally compact integration. The absence of distinct frequency degenerate counter-propagating modes in the FP geometry and the single mode nature of the PCRs avoids potential complications from unwanted AMXs and permits transferring concepts from fiber-ring cavities with optical isolators to integrated photonic chips (without isolators). As each unit cell of the PCRs can, in principle, be individually tailored, this opens a huge design space e.g. for dispersion tailoring and bandwidth extension through chirped reflectors. 
With relevance to quantum photonics, bio-chemical sensing and astronomical spectrograph calibration, the presented results provide a resonator platform that, through customized PCRs, may lead to DKSs at visible wavelengths, and at other wavelengths that are currently inaccessible due to strong normal material dispersion. 
Besides DKS and comb generation, the demonstrated resonator platform could also prove useful for integrated optical parametric oscillators \cite{bruch:2019} and optical harmonic generation \cite{wolf:2018}. Immediate further research opportunities leveraging the specific characteristics of the new resonators include dispersion managed solitons \cite{jang:2014, bao:2015, dong:2020}, \mbox{sinc-,} Nyquist- and zero-dispersion solitons \cite{turitsyn:2020, xue:2021, zhang:2022}, nonlinear ‘gain-through-loss’ \cite{perego:2018}, slow-light \cite{lu:2022}, spectral engineering \cite{lucas:2022, moille:2022} and filter-driven pulse formation \cite{bale:2008, dong:2021}, which bodes well for a new generations of integrated broadband and ultrafast light sources.

\subsection*{Methods}
\small
\paragraph{Sample fabrication.} 
The samples were fabricated commercially by LIGENTEC SA using UV optical lithography. When fabricating devices with features close to the resolution limit of optical lithography, deviations between designed and fabricated geometry can arise that, if unaccounted for, can significantly alter device performance. In the case of the PCRs presented in this work, this results in a shift in the mean waveguide width and a reduction of the effective corrugation depth of the cells. The parameterized geometry of the PCRs enables us to correct for this by establishing an interpolation table between designed and fabricated geometries which is used to preemptively adjust the mask in order to achieve the targeted final-geometry. Thus, through careful process calibration, it is possible to fabricate the microresonators in a wafer-level process using optical lithography without the need for e-beam lithography, which photonic crystal devices often require.
\paragraph{Resonator dispersion.} 
\newcommand\underrel[2]{\mathrel{\mathop{#2}\limits^{#1}}}
The introduction of a sub-wavelength periodic perturbation via the PCR couples the forward- and backward-propagating waves together through a coupling coefficient $\kappa$ which can be estimated from the half-bandgap opening $\delta \omega / 2$ divided by the group velocity of the unperturbed waveguide \cite{haus:1984}:
\begin{equation}
    |\kappa| \approx \frac{\delta\omega}{2} \left. \frac{d\beta}{d\omega} \right|_{\omega=\omega_0} 
\end{equation}
where $\omega_0$ is the bandgap center frequency such that \mbox{$\beta(\omega_0) = \pi / \Lambda $}, $\Lambda$ being the PCR period. Within the bandgap the eigenvalues of the coupled system are real, and the waves decay exponentially with rate $\gamma = - \sqrt{|\kappa|^2 - \delta^2}$ where $\delta = \beta(\omega) - \pi/\Lambda$ is the detuning parameter. By introducing the appropriate boundary conditions, the reflection coefficient can be obtained as a function of the PCR length $l$ :
\begin{equation}
    \Gamma = \frac{ \kappa^*  \cdot \sinh{\gamma l}}{\gamma \cdot \cosh{\gamma l} + j \delta \cdot \sinh{\gamma l}}  
\end{equation}
\begin{equation}
    \tan{\angle \Gamma} = \frac{\delta}{\gamma} \cdot \tanh{\gamma l} 
\end{equation}%
The round-trip phase of the FP cavity is then given by the sum of the PCR and waveguide contributions:
\begin{equation}
    \label{eq:roundtrip_phase}
    \phi = 2 L \beta + 2 \arctan \left( \frac{\delta}{\gamma} \cdot \tanh{\gamma l} \right)
\end{equation}
where $L$ is the one-way length of the intracavity waveguide.

The resonator roundtrip $\mathrm{GDD} = \partial^2\phi / \partial\omega^2$ therefor also contains the sum of the contributions from the PCRs (two reflections) and that of the waveguide between the PCRs (back and forth). It is related to the microresonator dispersion $D_2$ via $\mathrm{GDD}=-2\pi D_2/D_1^3$. For DKS generation we drive the resonator at a wavelength where the waveguide single pass contribution to the GDD is \SI{-445}{\fs^2}, and the contribution of the PCR from a single reflection is \SI{-2170}{\fs^2}. For both values, $\tau_0^2 \gg |\mathrm{GDD}|$ for a DKS duration of $\tau_0 \approx \SI{100}{\fs}$, i.e. the effect of a single reflection or a single pass through the waveguide has negligible impact on the pulse; therefore the effect of both contributions of the dispersion can be lumped together.

The fit in \reffig{fig:2}{d} is obtained by fitting the resonance frequencies against their mode number $\mu_0 + \mu = \phi/(2\pi)$ using eq.~\ref{eq:roundtrip_phase} and expanding the wavenumber as $\beta(\omega) = \beta_0 + \beta_1 (\omega - \omega_0) + \frac{1}{2}\beta_2(\omega - \omega_0)^2 $.

\paragraph{Soliton generation and soliton spectra.}
In order to tune into the respective solitons states without the requirement for a rapid tuning scheme (i.e. faster than the resonator thermal decay rate), we use the established auxiliary laser method \cite{zhang:2019}. In this scheme a secondary auxiliary laser (\SI{1560}{\nm}) coupled to the blue-side of a resonance of the TE mode family (\reffig{fig:1}{b}), of similar on-chip power (\SI{150}{\mW}) and orthogonal to the TM-polarized pump light, is used to thermally stabilize the resonator and conveniently mitigate thermal shifts occurring when tuning the main pump laser by keeping the total intracavity power nearly constant.

All soliton spectra show a modulated spectral envelope (\reffig{fig:3}{b}) with a period and amplitude of \SI{\sim 207}{\GHz} and \SI{\sim1.5}{\decibel} respectively. We attribute this to reflections between the chip facet and the PCR, forming a low-Q cavity in the connecting waveguide of length \SI{\sim380}{\um}.

\subsection*{Author Contributions}
T.W. designed the resonators, performed the experiments, analyzed the data and carried out numeric simulations. M.G. supported the design of the resonators. T.V. and M.L. supported the experiments. T.W. and T.H. wrote the manuscript. T.H. supervised the work. 

\subsection*{Funding}
\small
This project has received funding from the European Research Council (ERC) under the EU’s Horizon 2020 research and innovation program (grant agreement No 853564), from the EU’s Horizon 2020 research and innovation program (grant agreement No 965124)and through the Helmholtz Young Investigators Group VH-NG-1404; the work was supported through the Maxwell computational resources operated at DESY.

\subsection*{Disclosures}
\small All authors declare no conflict of interest.

\subsection*{Acknowledgements}
The authors acknowledge valuable discussions with \mbox{Alexander Yu. Petrov} regarding the analytic computation of the PCR's dispersion.

\printbibliography

\end{document}